\documentclass[twocolumn,aps,prl,showpacs,preprintnumbers,nofootinbib]{revtex4}
\usepackage{amsfonts,amssymb,amsmath}
\usepackage{graphicx}
\newcommand{\f}[2]{\frac{#1}{#2}}
\newcommand{\mk}[1]{\left( #1 \right)}
\newcommand{\kk}[1]{\left[ #1 \right]}

\newcommand{\be}{\begin{equation}}
\newcommand{\ee}{\end{equation}}
\newcommand{\lcdm}{{\rm \Lambda CDM}}
\def\eff{{\rm eff}}
\begin{document}
\preprint{RESCEU-5/12}

\title{
Cosmology Based on $f(R)$ Gravity Admits $1$~eV Sterile Neutrinos
}

\author{Hayato~Motohashi$^{~1,2}$}
\author{Alexei A.\ Starobinsky$^{~2,3}$}
\author{Jun'ichi Yokoyama$^{~2,4}$}
\address{
$^{1}$ Department of Physics, Graduate School of Science,
The University of Tokyo, Tokyo 113-0033, Japan \\
$^{2}$ Research Center for the Early Universe (RESCEU),
Graduate School of Science, The University of Tokyo, Tokyo 113-0033, Japan \\
$^{3}$ L. D. Landau Institute for Theoretical Physics RAS,
Moscow 119334, Russia \\
$^{4}$ Kavli Institute for the Physics and Mathematics of the Universe (Kavli IPMU),
The University of Tokyo, Kashiwa, Chiba 277-8568, Japan
}

\begin{abstract}
It is shown that the tension between recent neutrino oscillation experiments,
favoring sterile neutrinos with masses of the order of 1 eV, and cosmological
data which impose stringent constraints on neutrino masses from
the free streaming suppression of density fluctuations, can be resolved
in models of the present accelerated expansion of the Universe based on $f(R)$
gravity.
\end{abstract}
\maketitle

The possibility of the existence of light sterile neutrinos
has been suggested by neutrino oscillation experiments such as
LSND~\cite{Aguilar:2001ty} and
MiniBooNE~\cite{AguilarArevalo:2010wv}
as well as by the gallium anomaly of
the SAGE and
GALLEX experiments
(see \cite{Gavrin:2010qj} for a discussion and original references).
Recent nuclear reactor experiments also favor the additional neutrinos
with masses of this range~\cite{Mention:2011rk}.
Further analysis shows increasing experimental evidence that
there may exist one or two species of sterile neutrinos with masses
of the order of 1 eV \cite{Giunti:2011hn}. 

From the cosmological point of view, recent reanalysis of the
primordial helium abundance produced at the big bang nucleosynthesis
(BBN) also favors the existence of  extra components of
radiation~\cite{Izotov:2010ca}. In terms of
the effective number of neutrinos, which is defined by the
total energy density of the radiation as
\be \rho_r=\rho_\gamma\left[1+N_\eff\f{7}{8}\left(\f{4}{11}\right)^{4/3}\right] \ee
with $\rho_\gamma$ being the energy density of the photons, they find
$N_\eff=3.68^{+0.80}_{-0.70}~(2\sigma)$ or
$N_\eff=3.80^{+0.80}_{-0.70}~(2\sigma)$ for the neutron
lifetime $\tau_n=885.4\pm 0.9$s or $878.5\pm 0.8$s, respectively.
Note that the standard three flavor neutrino species gives
$N_\eff=3.046$~\cite{Mangano:2005cc}.
Furthermore, the cosmic microwave background (CMB) anisotropy
observations at the small angular scales yield similar
values, $N_\eff=4-5$~\cite{Komatsu:2010fb,Dunkley:2010ge}.
Thus both the BBN and the CMB suggest there exists extra relativistic species,
which may be sterile neutrinos that are expected to be thermalized in
the early Universe due to mixing~\cite{Kainulainen:1990ds}.

If we further incorporate
the large-scale-structure (LSS) data in the cosmological analysis,
however, it turns out that  in the standard
flat ${\rm \Lambda}$-cold-dark-matter ($\lcdm$) model
the sterile neutrino mass is constrained to be appreciably smaller
 than $1$~eV~\cite{Hamann:2010bk,Burenin:2012uy}
to avoid suppression of small-scale fluctuations due to free streaming.

Hence apparently there exists a tension between the experimental data of
neutrino oscillations and the LSS data. But of course they should
not be treated on an equal footing because the latter requires a number of
assumptions about the cosmic evolution. Indeed there have been some
attempts to make cosmology compatible with these experimental 
data~\cite{Hamann:2011ge} by adopting a $w$CDM model to
treat the equation-of-state parameter of the dark energy, $w$, as an
additional fitting parameter, or by introducing extra radiation
components besides the sterile neutrinos. However, the former results
in $w<-1$ and a larger CDM abundance with the cosmic age being significantly
smaller than the standard value, while the latter solution may be
in conflict with the aforementioned constraint from the BBN. So neither
is an attractive solution.

In this {\em Letter}, we show that the extra growth of small-scale fluctuations
at recent redshifts which occurs in viable cosmological models of the present
accelerated expansion of the Universe (in other terms, in models of the present
dark energy) based on $f(R)$
gravity~\cite{Hu:2007nk,Appleby:2007vb,Starobinsky:2007hu} can make eV-mass
sterile neutrinos compatible with cosmological observations under the proper
choice of the function $f$. $f(R)$ gravity is a simple generalization of
General Relativity (GR) obtained by introducing a phenomenological function
of the Ricci curvature $R$, see {\it e.g.}, the recent review~\cite{DT2010}. It
represents a special case of more general
scalar-tensor gravity with the Brans-Dicke parameter $\omega_{BD}=0$, and
it has an extra scalar degree of freedom (or, scalar particle dubbed a scalaron). 
However, in contrast to Brans-Dicke gravity, the scalaron is
massive and its rest mass $M_s$ depends on $R$ , {\it i.e.}, on the background matter
density in the regime of small deviations from GR. Such models explain the
present cosmic acceleration without introducing a cosmological constant;
mathematically this means that $f(0)=0$. For the $f(R)$ models of the present dark
energy constructed in \cite{Hu:2007nk,Appleby:2007vb,Starobinsky:2007hu} which
satisfy all existing observational data, the deviation of the background
evolution from the standard $\lcdm$ model is small, less than a few percent
(see {\it e.g.}, \cite{Motohashi:2010tb}), though not exactly zero. Their most dramatic
difference from the standard model appears in the enhancement of the
evolution of matter density perturbations on scales smaller than
the Compton wavelength of the scalaron field that occurs at redshifts
of the order of a few depending on the scale. This extra growth can compensate
the suppression due to the free streaming of massive neutrinos and thus the
upper bound for the neutrino mass is relaxed in $f(R)$ gravity~\cite{Motohashi:2010sj}.
We show that the same mechanism works in the case of sterile neutrinos, too, to
make cosmology with them compatible with neutrino experiments.

$f(R)$ gravity is defined by the action
\be \label{ac} S=\f{1}{16\pi G} \int d^4x \sqrt{-g} f(R) +S_m, \ee
where $S_m$ is the action of the matter content which is assumed
to be minimally coupled to gravity (we put $\hbar=c=1$). If we set $f(R)=R-2\Lambda$,
it reproduces GR with a cosmological constant. Instead, for definiteness we use the
following form \cite{Starobinsky:2007hu}
\be \label{fr} f(R)=R + \lambda R_s \kk{\mk{1+\f{R^2}{R_s^2}}^{-n}-1}, \ee
where $n,~\lambda$, and $R_s$ are model parameters.
Two of them are free parameters and the other one is determined by the
other two and observational data. If we take $n$ and $\lambda$ as
free parameters, $R_s$ is approximately proportional to
$\lambda^{-1}$~\cite{Motohashi:2010tb}. Note that $n$ should be
taken sufficiently large, $n \gtrsim 2$, if we want to obtain a noticeable
effect for the density perturbation enhancement (see below).
The model \eqref{fr} can describe the accelerated
expansion of the present Universe and it quickly approaches the $\lcdm$ model
for redshift $z>1$ if we take large $n$ and $\lambda$. Strictly speaking, a term proportional to
$R^2$ should be added to (\ref{fr}) to avoid the scalaron mass
$M_s$ exceeding the Planck mass for high, but not too high matter densities
in the early Universe~\cite{Starobinsky:2007hu}, as well as to exclude the
possible formation of an extra weak curvature singularity in the recent past
which was found in~\cite{F2008,AB2008} (still there remains an open question as to
what would occur instead of this singularity~\cite{AD2011}). However, the 
coefficient of this term [usually written as $(6M^2)^{-1}$ where $M$ coincides
with the scalaron mass $M_s$ in the regime when this term dominates other non-GR
terms in (\ref{fr})] should be very small in order not to destroy the standard
evolution of the early Universe. Namely, either $M$ should be larger than
the Hubble parameter $H$ at the end of inflation, or this term can drive inflation
by itself if $M\approx 3\times 10^{13}$ GeV~\cite{Starobinsky:1980te,Appleby:2009uf}.
Thus, the $R^2$ correction is negligibly small at present curvatures.

The model \eqref{fr} describes a similar background expansion history to that of the $\lcdm$ model. Although the equation-of-state parameter for dark energy makes a phantom crossing at $z\sim 3$~\cite{Hu:2007nk,Motohashi:2010tb,Motohashi:2011wy}, it does not change the CMB spectrum significantly.   
On the other hand, the fluctuations evolve differently. We define the metric perturbation by the following notation,
\be ds^2=-(1+2\Phi)dt^2+a^2(t)(1-2\Psi)\delta_{ij}dx^idx^j. \ee
We can derive the effective gravitational constant and the gravitational slip in $f(R)$ gravity by using the subhorizon limit and the quasistatic approximation~\cite{Song:2006ej}:
\be
\f{k^2}{a^2}\Phi=-4\pi G_{\rm eff}(t,k)\rho \Delta, \quad \f{\Psi}{\Phi}=\eta(t,k)\\
\ee
where $k$ is the comoving wave number, and
\begin{align}
\label{Geff} \f{G_{\rm
 eff}(t,k)}{G}=\f{1}{f'}\f{1+4\f{k^2}{a^2}\f{f''}{f'}}{1+3\f{k^2}{a^2}\f{f''}{f'}},
 ~~
\eta(t,k)=\f{1+2\f{k^2}{a^2}\f{f''}{f'}}{1+4\f{k^2}{a^2}\f{f''}{f'}}.
\end{align}
Here, $\Delta$ is the gauge-invariant comoving matter perturbation, and the prime denotes the derivative with respect to the Ricci curvature. Thus, in the quasi-GR regime when $f'\approx 1$, the effective gravitational
constant can become up to $33\%$ larger, independently of a detailed functional form of $f(R)$. 
This is the cause of the enhancement of perturbation growth.

As a result of the time and the scale dependences of these parameters, evolution of matter density fluctuations is different from that in the $\lcdm$ model; namely, it is enhanced on small scales~\cite{Hu:2007nk,Starobinsky:2007hu,Motohashi:2009qn,Motohashi:2010tb}.
On the contrary, the light neutrinos suppress structure formation by free streaming. Therefore, $f(R)$ modification and neutrino masses play opposite roles in the growth of perturbations and thus the allowed range for the total neutrino mass is relaxed in $f(R)$ gravity, compared with the $\lcdm$ model~\cite{Motohashi:2010sj}. We can apply this mechanism for the case of sterile neutrinos.

Needless to say, the existence and the mass of the sterile neutrino is to be determined by ground-based experiments rather than cosmology which also depends on other factors that fix the evolution of the background and perturbed Universe in a complicated manner.

Therefore, the question we address in the present {\em Letter} is as follows: If future experiments fix the sterile neutrino mass at the order of $1$~eV,
that is inconsistent with the standard $\lcdm$ cosmology
(see below), can the earlier proposed cosmological models in the scope
of $f(R)$  gravity save the situation and make cosmology compatible with 
particle physics?
We carried out a Markov chain Monte Carlo (MCMC) analysis for the $\lcdm$ model and $f(R)$ gravity with one sterile neutrino with a mass of $1$~eV. 
Practically, we have neglected the rest masses of three standard neutrino species
(assuming that $\sum_{i=1}^3m_{\nu i} < 0.1$ eV) compared to that of the
sterile neutrino. We have modified the
{\sc mgcamb} code~\cite{Lewis:1999bs}, which provides the evolution
of the modified growth of matter fluctuations by setting functional forms
of $G_{\rm eff}(t,k)$ and $\eta(t,k)$, so that it can implement $f(R)$
gravity by adopting \eqref{Geff}. 
We have not changed the background evolution equations, {\it i.e.}, we kept those in the $\lcdm$ model, because the difference between the background evolution in the viable $f(R)$ model and the $\lcdm$ model is not significant (though it is not exactly zero).  We fixed the contribution from the $1$ eV sterile neutrino by using the relation 
$ \Omega_\nu h^2={\sum m_\nu}/{94.1 {\rm eV}}$. 
We plugged the above modified {\sc mgcamb} code into {\sc cosmomc}~\cite{Lewis:2002ah,cosmomc} 
for our MCMC analysis to search for the best fit set of the model parameters.
The free parameters are the density parameters for the dark matter $\Omega_{\rm DM}h^2$ and for the baryon $\Omega_{\rm b}h^2$, the sound horizon angle 
$\theta_*\equiv 100r_s(z_*)/D_A(z_*)$, 
whose use is helpful to minimize degeneracies among the cosmological parameters~\cite{cosmomc},
the optical depth $\tau$, the scalar spectral index $n_s$, the amplitude of the primordial power spectrum $\ln(10^{10}A_s)$, the Sunyaev-Zel'dovich template normalization $A_{\rm SZ}$, 
the linear galaxy bias parameter $b_0$, 
and, for the case of $f(R)$ gravity, the amplitude of the $f(R)$ modification $\lambda$. Here, $\Omega_{\rm DM}$ means the sum of the contribution from cold dark matter and massive neutrinos. 

\begin{figure}[t]
\centering
\includegraphics[width=53mm]{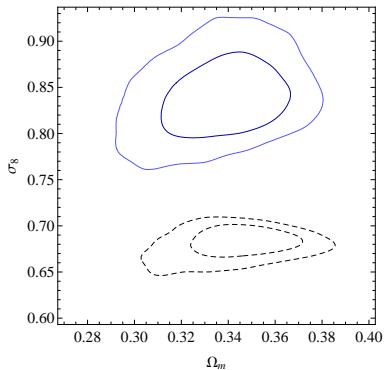}
\caption{
$1\sigma$ and $2\sigma$ contours of $\sigma_8$ for the cases with three massless and one massive neutrino with the mass being $1$ eV in the $\lcdm$ model (dashed black) and $f(R)$ gravity (solid blue).
}
\label{fig:sigma8}
\end{figure}

In contrast to $\lambda$, which is treated as a floating parameter, we have fixed another $f(R)$ parameter $n$ to $2$. This value is chosen because it is the minimal integer value for which the scalaron mass $M_s$
given by $M_s^2=1/3f''(R)$ in the quasi-GR regime is, on one hand, not much higher than the Hubble constant $H_0$ if estimated at
the present cosmic background matter density $\rho_{m0}=3\Omega_mH_0^2/(8\pi G)$ (see the value of the constant $B_0$ below which characterizes it quantitatively), and on the other hand, it is already sufficiently large for matter densities inside the Solar System (not speaking about those in laboratory experiments) to make the scalaron heavy and unobservable even outside gravitating bodies similar to the dilaton in string theory. Indeed, for the functional form (\ref{fr}), $M_s\propto \rho_m^{n+1}$ in the
quasi-GR regime for $\rho_m\gg \rho_{m0}$. Thus, here there is no necessity to consider the more subtle chameleon effect which
can make scalaron unobservable even if it is light outside bodies (though heavy inside them).
Further, in order to have the future stable de Sitter stage, $\lambda$ should be larger than $0.95$.
For $n=2$ and $\lambda=0.95$, the deviation index $B_0=(f''/f')(dR/d\ln H)|_{t=t_0}$ is not too small nor too large, namely, $B_0=0.21$~\cite{Motohashi:2010tb}. This value is in agreement with the upper limit $B_0<0.4$ recently obtained in~\cite{GMSM2010}. 
On the contrary, the much more stringent upper limit obtained from cluster abundance in~\cite{SVW2009}
does not apply to our model because it was obtained for a similar functional form of $f(R)$ introduced in~\cite{Hu:2007nk},
with its parameter value characterizing the large-$R$ behaviour corresponding to $n=0.5$ in (\ref{fr}).

To constrain the free parameters, we used the observational data of CMB by WMAP7~\cite{Komatsu:2010fb} and a power spectrum of luminous red galaxies (LRG) by SDSS DR7~\cite{Reid:2009xm} within the wave number range $0.02~h{\rm Mpc}^{-1} \leq k \leq 0.08~h{\rm Mpc}^{-1}$ to avoid the ambiguities in nonlinear evolution. Based on \cite{Reid:2009xm}, we computed $P_{\rm halo}(k)$ to compare the observational power spectrum of the LRG sample. We floated the galaxy bias $b_0$ in (15) in \cite{Reid:2009xm} assuming that $b_0$ takes the same value for the LRG sample, while $a_1$ and $a_2$ are analytically marginalized.

As a result, we find that in the presence of a $1$~eV massive sterile neutrino $f(R)$ gravity fits the cosmological data much better than the $\lcdm$ model. In terms of the best-fit $\chi_\eff^2$ value, $\chi_{\lcdm}^2-\chi_{{\rm fRG}}^2=9.55$.
According to the Akaike information criteria (AIC)~\cite{Akaike}, if $\chi_\eff^2$ improves by $2$ or more with a new additional fitting parameter, its incorporation is justified. In this context, the performance of $f(R)$ gravity, which includes only one more additional parameter beyond the $\lcdm$ model, improves $\chi_\eff^2$ very well.

The best-fit value for the $f(R)$ parameter is $\lambda=7.41$. The relatively large value of $\lambda$ leads practically to the same background evolution as that in the $\lcdm$ model. Thus, it is important to focus on quantities referring to perturbations.
Indeed, comparing the best-fit parameter values in the two models, 
most are similar but significant difference shows up in $b_0$ and $\sigma_8$.

The value of the galaxy bias parameter $b_0$ helps to understand the physics. While the $\lcdm$ model takes $b_0=1.26 \pm 0.11$, $f(R)$ gravity gives $b_0=1.16 ^{+0.14}_{-0.23}$. The general enhancement of matter fluctuations in $f(R)$ gravity slightly reduces the galaxy bias by $\sim 0.1$.

Of course, the enhancement of matter fluctuations in $f(R)$ gravity is scale dependent and operates on small scales only. Therefore, it is different from the scale-independent galaxy bias. Figure \ref{fig:sigma8} depicts the $1\sigma$ and $2\sigma$ contours for the particular, but very important for observations, quantity $\sigma_8$ in the $\lcdm$ model and $f(R)$ gravity. 
The $\lcdm$ model with a $1$~eV massive neutrino yields $\sigma_8=0.692_{-0.043}^{+0.020}$, 
which is in disagreement with observations~\cite{Burenin:2012uy,Reid:2009xm}. $f(R)$ gravity, on the contrary, works well in this respect with $\sigma_8=0.838_{-0.08}^{+0.13}$.

\begin{figure}[t]
\centering  
\includegraphics[width=65mm]{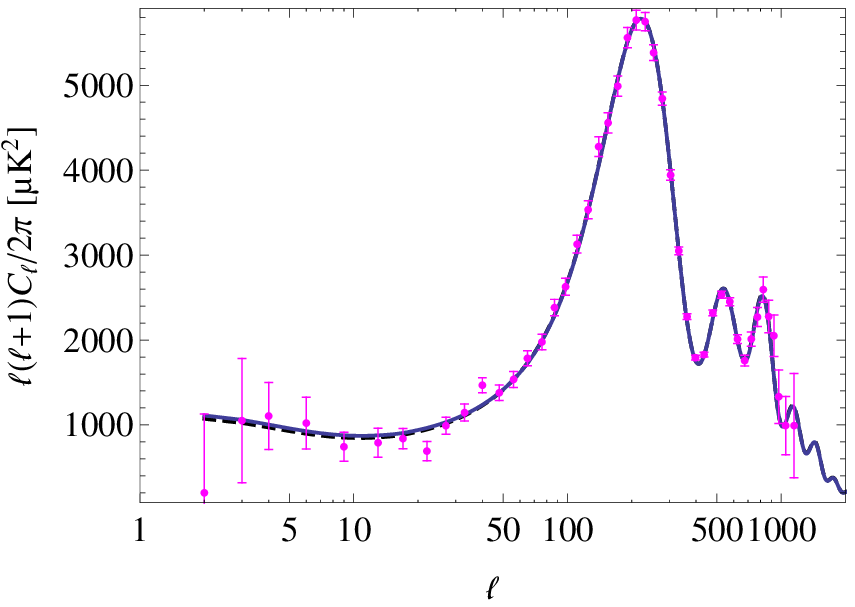}
\includegraphics[width=65mm]{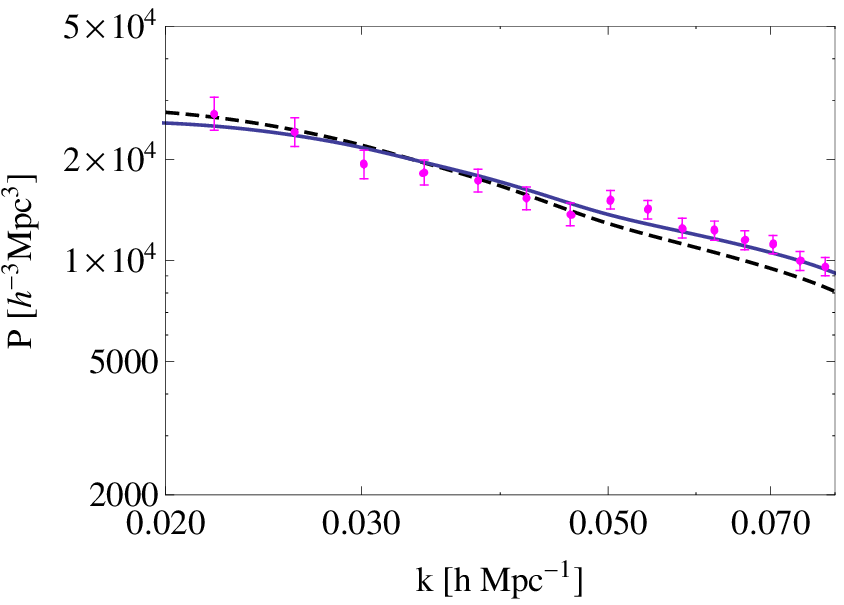}
\caption{
CMB temperature power spectrum with WMAP7 data (upper panel) and galaxy power spectrum $b_0^2P_{\rm DM}(k)$
at $z=0.2$ with SDSS data (lower panel). The lines show the best-fitted $f(R)$ gravity (solid blue) and the best-fitted $\lcdm$ model (dashed black) in the presence of three massless and a $1$~eV massive neutrino. The difference of the best-fit $\chi^2$ between them is $\chi_{\lcdm}^2-\chi_{{\rm fRG}}^2=9.55$.
}
\label{fig:clpk}
\end{figure}

More clearly, we see that $f(R)$ gravity fits the cosmological data much
better from Fig.~\ref{fig:clpk}, in which we present the CMB temperature
power spectrum and the galaxy power spectrum for the best-fit parameters in
the $\lcdm$ model and $f(R)$ gravity. As expected, the CMB anisotropy is
similar in both models but the galaxy power spectrum is remarkably
different. To see the differences between the $\lcdm$ model and $f(R)$
gravity directly, we show galaxy power spectrum $b_0^2 P_{\rm DM}(k)$ at
$z=0.2$ in the lower panel of Fig.~\ref{fig:clpk}, 
since for this wave number range one finds $P_{\rm halo}\simeq b_0^2 P_{\rm DM}(k)$.
The tilt of the power spectrum is different because 
$f(R)$ gravity efficiently counteracts the suppression by the free
streaming of $1$~eV sterile neutrinos. 
This fact leads to $f(R)$ gravity fitting the observational data 
significantly better than the standard $\lcdm$ model.

Thus we conclude that if an $\sim 1$~eV sterile neutrino is indeed established by ground-based experiments, that is in fact favored by a number of experiments now, then cosmology based on $f(R)$ gravity removes the problem with the
inadmissibly low value of $\sigma_8$ and fits the CMB and galaxy data much 
better than the standard $\lcdm$ cosmology.

H.M.\ and J.Y.\ thank M.~Nakashima, K.\ Ichiki, S.~Saito and A.~Oka 
for helpful discussions and technical support.  
A.S.\ acknowledges RESCEU hospitality as a visiting professor.  
He was also partially supported by RFBR Grant No. 11-02-00643 and by the  
Scientific Programme ``Astronomy'' of the Russian Academy of  
Sciences. This work was partially supported  by JSPS Research  
Fellowships for Young Scientists (H.M.), JSPS Grant-in-Aid for  
Scientific Research No.\ 23340058 (J.Y.), 
and JSPS Grant-in-Aid for Scientific Research on Innovative Areas No. 21111006 (J.Y.).


\begin{thebibliography}{99}
\bibitem{Aguilar:2001ty}   
  A.~Aguilar-Arevalo {\it et al.}  [LSND Collaboration],  
  Phys.\ Rev.\ D {\bf 64}, 112007 (2001).  
\bibitem{AguilarArevalo:2010wv}   
  A.~A.~Aguilar-Arevalo {\it et al.}  [The MiniBooNE Collaboration],  
  Phys.\ Rev.\ Lett.\  {\bf 105}, 181801 (2010).  
\bibitem{Gavrin:2010qj} 
  V.~N.~Gavrin, V.~V.~Gorbachev, E.~P.~Veretenkin and B.~T.~Cleveland,
  arXiv:1006.2103.
\bibitem{Mention:2011rk}   
  G.~Mention {\it et al.}, 
  Phys.\ Rev.\ D {\bf 83}, 073006 (2011).  
\bibitem{Giunti:2011hn}   
  C.~Giunti and M.~Laveder,  
  Phys.\ Rev.\ D {\bf 84}, 093006 (2011);  
   {\bf 84}, 073008 (2011); 
  J.~Kopp, M.~Maltoni, and T.~Schwetz,  
  Phys.\ Rev.\ Lett.\  {\bf 107}, 091801 (2011).  
\bibitem{Izotov:2010ca}   
  Y.~I.~Izotov and T.~X.~Thuan,  
  Astrophys.\ J.\  {\bf 710}, L67 (2010).  
\bibitem{Mangano:2005cc}   
  G.~Mangano {\it et al.}, 
  Nucl.\ Phys.\ B {\bf 729}, 221 (2005).  
\bibitem{Komatsu:2010fb}   
  E.~Komatsu {\it et al.}  [WMAP Collaboration],  
  Astrophys.\ J.\ Suppl.\  {\bf 192}, 18 (2011).  
\bibitem{Dunkley:2010ge}   
  J.~Dunkley 
 {\it et al.},  
  Astrophys.\ J.\  {\bf 739}, 52 (2011);
  R.~Keisler 
 {\it et al.}, 
{\it ibid.} {\bf 743}, 28 (2011);   
  M.~Archidiacono, E.~Calabrese, and A.~Melchiorri,  
  Phys.\ Rev.\ D {\bf 84}, 123008 (2011).  
\bibitem{Kainulainen:1990ds}   
  K.~Kainulainen,  
  Phys.\ Lett.\ B {\bf 244}, 191 (1990). 
\bibitem{Hamann:2010bk}   
  J.~Hamann, S.~Hannestad, G.~G.~Raffelt, I.~Tamborra, and Y.~Y.~Y.~Wong,  
  Phys.\ Rev.\ Lett.\  {\bf 105}, 181301 (2010);  
  E.~Giusarma, M.~Corsi, M.~Archidiacono, R.~de Putter, A.~Melchiorri, O.~Mena, and S.~Pandolfi,  
  Phys.\ Rev.\ D {\bf 83}, 115023 (2011). 
\bibitem{Burenin:2012uy}   
  R.~A.~Burenin and A.~A.~Vikhlinin,  
Astron. Lett. {\bf 38}, 347 (2012).
\bibitem{Hamann:2011ge}   
  J.~Hamann, S.~Hannestad, G.~G.~Raffelt and Y.~Y.~Y.~Wong,  
  JCAP {\bf 1109}, 034 (2011);  
  J.~R.~Kristiansen and O.~Elgaroy,  
  arXiv:1104.0704.  
\bibitem{Hu:2007nk}  
  W.~Hu and I.~Sawicki,  
  Phys.\ Rev.\  D {\bf 76}, 064004 (2007).  
\bibitem{Appleby:2007vb}  
  S.~A.~Appleby and R.~A.~Battye,  
  Phys.\ Lett.\  B {\bf 654}, 7 (2007).  
\bibitem{Starobinsky:2007hu}  
  A.~A.~Starobinsky,  
  JETP Lett.\  {\bf 86}, 157 (2007).  
\bibitem{DT2010}  
  A.~De~Felice and S.~Tsujikawa,  
  Living Rev.\ Rel.\ {\bf 13}, 3 (2010).  
\bibitem{Motohashi:2010tb}  
  H.~Motohashi, A.~A.~Starobinsky, and J.~Yokoyama,  
  Prog.\ Theor.\ Phys.\  {\bf 123}, 887 (2010).  
\bibitem{Motohashi:2010sj}  
  H.~Motohashi, A.~A.~Starobinsky, and J.~Yokoyama,  
  Prog.\ Theor.\ Phys.\  {\bf 124}, 541 (2010).  
\bibitem{F2008}  
  A.~V.~Frolov,  
  Phys.\ Rev.\ Lett.\  {\bf 101}, 061103 (2008).   
\bibitem{AB2008}  
  S.~A.~Appleby and R.~A.~Battye,
  JCAP {\bf 0805}, 019 (2008).
\bibitem{AD2011}  
  E.~V.~Arbuzova and A.~D.~Dolgov,  
  Phys.\ Lett.\ B {\bf 700}, 289 (2011).  
\bibitem{Starobinsky:1980te}  
  A.~A.~Starobinsky,  
  Phys.\ Lett.\  B {\bf 91}, 99 (1980).  
\bibitem{Appleby:2009uf}   
  S.~A.~Appleby, R.~A.~Battye and A.~A.~Starobinsky,  
  JCAP {\bf 1006}, 005 (2010).  
\bibitem{Motohashi:2011wy}  
  H.~Motohashi, A.~A.~Starobinsky, and J.~Yokoyama,  
  JCAP {\bf 1106}, 006 (2011).  
\bibitem{Song:2006ej} 
  Y.~-S.~Song, W.~Hu and I.~Sawicki,
  Phys.\ Rev.\ D {\bf 75}, 044004 (2007);
  S.~Tsujikawa,
{\it ibid.} {\bf 76}, 023514 (2007).
\bibitem{Motohashi:2009qn}  
  H.~Motohashi, A.~A.~Starobinsky, and J.~Yokoyama,  
  Int.\ J.\ Mod.\ Phys.\  {\bf D18}, 1731 (2009).  
\bibitem{Lewis:1999bs}  
  A.~Lewis, A.~Challinor, and A.~Lasenby,  
  Astrophys.\ J.\  {\bf 538}, 473 (2000);  
  A.~Hojjati, L.~Pogosian and G.-B.~Zhao,  
  JCAP {\bf 1108}, 005 (2011).  
\bibitem{Lewis:2002ah}   
  A.~Lewis and S.~Bridle,  
  Phys.\ Rev.\ D {\bf 66}, 103511 (2002).  
\bibitem{cosmomc}  
  {\sc cosmomc}, http://cosmologist.info/cosmomc.  
\bibitem{GMSM2010}  
  T.~Giannantonio, M.~Martinelli, A.~Silvestri, and A.~Melchiorri,  
  JCAP {\bf 1004}, 030 (2010).  
\bibitem{SVW2009}   
  F.~Schmidt, A.~Vikhlinin and W.~Hu,   
  Phys.\ Rev.\ D {\bf 80}, 083505 (2009).  
\bibitem{Reid:2009xm}  
  B.~A.~Reid 
 {\it et al.},  
  Mon.\ Not.\ Roy.\ Astron.\ Soc.\  {\bf 404}, 60 (2010).  
\bibitem{Akaike}  
  H. Akaike, IEEE Trans. Auto. Control {\bf 19}, 716 (1974).  
\end{thebibliography}
\end{document}